# Measurement Symmetry and Heisenberg Geometry: Embedding Classical Test Theory in a Noncommutative Representation

William R. Nugent, Ph.D.

https://orcid.org/0000-0003-0811-3221

College of Social Work, University of Tennessee, Knoxville, TN

Stokely Management Center

865-974-6544

wnugent@utk.edu

**Abstract**

Classical measurement theory is traditionally formulated in an algebraic framework. However, it is fundamentally commutative and does not naturally represent noncommutative measurement phenomena identified in the social sciences. This study investigates whether the transformation structure of classical measurement theory preserves a canonical noncommutative geometry. A measurement state vector is introduced, and transformations relating various forms of equivalence (parallelism) between measures are expressed as a Lie matrix group. A faithful matrix representation of the Heisenberg group is introduced, and the conjugation of Heisenberg elements by a general measurement transformation is derived. Results show this conjugation defines an automorphism of the Heisenberg group, preserving its commutator structure. However, if the elements of the measurement state vector are equally scaled the Heisenberg geometry is preserved exactly. The findings establish a symmetry linking classical measurement transformations with the Heisenberg group providing a mathematical foundation for extending classical measurement theory to phenomena exhibiting noncommutative structures.

**Measurement Symmetry and Heisenberg Geometry: Embedding Classical Test Theory in a Noncommutative Representation**

Classical measurement theory is built on a simple decomposition: an observed score is the sum of a true score and an error score, with zero covariance between them. This framework has anchored psychometrics since the early twentieth century, providing variance- and covariance-based definitions of reliability and a family of indices that remain central to modern measurement theory (Traub, 1997). Classical measurement theory is fundamentally a first- and second-moment theory, being based on means, variances, covariances, variance decompositions, and linear combinations. Consequently, it implicitly admits representation within a linear and commutative geometric framework (Allen & Yen, 1979; Atmanspacher & Romer, 2018; Lord & Novick, 1968; McDonald, 1999; Morgan, 2022).

Recent attention has begun to focus on noncommutative phenomena in the social, behavioral, and psychological sciences (Atmanspacher & Romer, 2018; Borsboom, 2005; Busemeyer & Bruza, 2024; Conte et al., 2008; Khrennikov, 2010; Schwarz & Sudman, 1992; Trueblood & Busemeyer, 2011; Wang et al., 2014). Noncommutative structures appear in several areas of social, behavioral, and psychological science, especially where the order or context of measurement changes the outcome. Robust order effects observed in surveys and decision making demonstrate that responses to item A followed by item B differ from responses to B followed by A, a signature of noncommuting cognitive observables (Busemeyer & Bruza, 2024; Wang et al., 2014). Research on contextuality in perception and judgment further shows that cognitive outcomes cannot be assigned a single joint distribution independent of context, aligning these effects with noncommutative probability rather than classical models (Atmanspacher & Filk, 2010; Dzhafarov et al., 2016). Several authors have called for mathematical and measurement frameworks capable of representing noncommutative phenomena (e.g., Atmanspacher & Römer 2018; Busemeyer & Bruza 2012; Dzhafarov et al., 2016; Khrennikov, 2010; Pothos & Busemeyer 2013; Wang et al. 2014). In response to these calls, the purpose of the present study is to investigate whether a transformation group arising from classical measurement theory possesses a symmetry that preserves a canonical noncommutative geometry, thereby providing a Lie-group geometric framework for representing noncommutative measurement transformations. Specifically, we ask whether classical measurement transformations preserve the intrinsic geometry of the Heisenberg group under conjugation.

Studying how scores on one measure transform into scores on another provides insight into their underlying structure and geometry (Golan, 2012; Krantz et al., 1971; Lord & Novick, 1968; Savov, 2017). The transformation rules in classical measurement theory that relate the traditional equivalence classes of measurement—parallel, tau-equivalent, essentially tau-equivalent, and congeneric—possess an underlying Lie-group structure that has received little attention. These transformations can be represented by upper-triangular matrices whose composition laws mirror those of a Lie matrix group (Stillwell, 2000), as explored recently by Nugent (2024, 2026). By studying how these classical measurement transformations act on a noncommutative geometry we can examine the degree to which these transformations preserve a noncommutative measurement structure. The results would have implications for the use of classical measurement theory for representing noncommutative phenomena in the social, behavioral, psychological, and related sciences.

These ideas are developed in this article by conceptualizing classical measurement theory within the sub-Riemannian geometry of the Heisenberg group. The choice of the Heisenberg group is motivated by both mathematical and measurement-theoretic considerations. Classical measurement transformations are naturally represented by upper-triangular affine matrices acting on measurement-state vectors, making it natural to ask how the simplest noncommutative upper-triangular Lie group behaves under these transformations. The Heisenberg group is the canonical step-2 nilpotent Lie group with this property. Its matrix representation naturally incorporates shifts, scalings, and interaction terms while remaining algebraically tractable. Moreover, its central coordinate is not introduced ad hoc; it is generated by the commutator relation $[X, Y] = Z$, providing a principled mathematical representation for interaction effects that may later be interpreted as accumulated measurement bias, order effects, or path dependence. These properties make the Heisenberg group a natural geometric extension of classical affine measurement transformations and motivate the conjugation analysis developed below. It will be shown that a particular class of measurement transformations characterized by equal scaling across components, denoted Γsym(γ,ω), acts by automorphism on a faithful representation of the Heisenberg group H(a,b,c). Under an equal-scaling symmetry, the conjugation action preserves the Heisenberg representation, and the commutator between the transformation group and the Heisenberg representation vanishes.

The next sections therefore provide a concise overview of classical measurement theory, emphasizing the decomposition of observed scores, the variance-based definition of reliability, and the linear transformations that relate scores from different measurement scales. The Heisenberg group is introduced and its geometry described (Binz & Pods, 2008; Capogna et al., 2007; Folland, 2025). The Heisenberg group is then conjugated by an affine transformation satisfying an equal-scaling symmetry. It will be shown that there exists a transformation group $\Gamma_{sym}(\gamma, \omega)$ that centralizes the Heisenberg group, producing preservation of the Heisenberg geometry. This central result of the analysis is that measurement transformations divide naturally into two classes. Under an equal-scaling symmetry the induced conjugation action preserves the Heisenberg structure, the commutator vanishes, and the subgroup generated by the Heisenberg group and the transformation group becomes a direct product. Consequently, the scaling symmetry identifies the precise condition under which a classical measurement transformation acts compatibly with a canonical noncommutative geometry. In this case the Heisenberg representation is preserved exactly. When the scaling symmetry is broken, the commutator becomes nontrivial and the representation is distorted. Thus, the uniform scaling symmetry identifies a distinguished measurement condition under which classical measurement transformations preserve an underlying noncommutative geometry.

## Overview of Classical Measurement Theory

Classical measurement theory is variance- and covariance based (Allen & Yen, 1979; Ghiselli et al., 1981; Haertel, 2006; Lord & Novick, 1968). Classical measurement theory begins with the decomposition of an observed score into a true component and an error component, and it develops the familiar variance-based definitions of reliability and measurement error from this foundation. Although traditionally presented in algebraic terms, this framework already contains structural elements that will later be reinterpreted geometrically. The decomposition $Y = \tau + \epsilon$ identifies two fundamental sources of variation, while the variance ratio defining reliability quantifies their relative contribution. Likewise, the equivalence classes, or forms of parallelism, of measurement instruments—parallel, tau-equivalent, essentially tau-equivalent, and congeneric—are characterized by linear transformations that relate their true-score and error components (Lord & Novick, 1968).

In classical measurement theory, an observed score for person $i$, on a measure A, $Y_{Ai}$, is assumed to consist of two components: a true score, $\tau_{Ai}$, and an error score, $\epsilon_{Ai}$:

$$Y_{Ai} = \tau_{Ai} + \epsilon_{Ai}.$$

In classical test theory, the true score is defined as the expected value, $E(Y_{Ai})$, of the observed score,

$$\tau_{Ai} = E(Y_{Ai}),$$

where the expectation is taken over hypothetical repeated measurements for person $i$, and the error score is defined as the deviation of the observed score from its expected value,

$$\epsilon_{Ai} = Y_{Ai} - \tau_{Ai} = Y_{Ai} - E(Y_{Ai})$$

(Allen & Yen, 1979; Lord & Novick, 1968; McDonald, 1999). By definition, true scores and error scores are uncorrelated, $\mathrm{Cov}(\tau_{A_i}, \epsilon_{A_i}) = 0$. Thus, the variance of observed scores, $\sigma^2(Y_{Ai})$, can be decomposed as,

$$\sigma^2(Y_{Ai}) = \sigma^2(\tau_{Ai}) + \sigma^2(\epsilon_{Ai}),$$

where $\sigma^2(\tau_{Ai})$ is the true-score variance and $\sigma^2(\epsilon_{Ai})$ is the error variance. The reliability of observed scores, $\mathrm{rel}(Y_A)$, is defined as the squared correlation between observed and true scores, $\rho^2(Y_A, \tau_A)$, or, equivalently, as the proportion of observed-score variance attributable to true-score variance,

$$\mathrm{rel}(Y_A) = \rho^2(Y_A, \tau_A) = \frac{\sigma^2(\tau_{Ai})}{\sigma^2(\tau_{Ai}) + \sigma^2(\epsilon_{Ai})} = \frac{\sigma^2(\tau_{Ai})}{\sigma^2(Y_{Ai})},$$

(Guttman, 1945; McDonald, 1999).

Classical measurement theory also defines several forms of forms of parallelism (or equivalence), of scores among measurement instruments, including parallel, tau-equivalent, essentially tau-equivalent, and congeneric measures (Haertel, 2006; Lord & Novick, 1968). Parallel measures have equal true scores and equal error variances. Tau-equivalent measures have equal true scores but may differ in error variances. Essentially tau-equivalent measures allow additive shifts in true scores while preserving true-score variance, though error variance may or may not differ.

Congeneric measures provide the most general case, allowing linear transformations of true scores and potentially different error variances (Haertel, 2006; Jöreskog, 1971; Lord & Novick, 1968; McDonald, 1999). For congeneric measures $A$ and $B$, the relationship between true scores is,

$$\tau_{B_i} = \gamma\tau_{Ai} + \omega, \quad (1)$$

where $\gamma > 0$ is a scaling constant and ω a real number ($\mathbb{R}$) additive constant. The congeneric model is especially important because it introduces the affine transformation structure that underlies the measurement transformation group developed later. As a consequence of eqn. (1), the true score SDs for measures A and B will be related by,

$$\sigma(\tau_B) = \gamma\sigma(\tau_A) \quad (2)$$

(Allen & Yen, 1979; Haertel, 2006; Jöreskog, 1971; Lord & Novick, 1968). Equation (2) shows that the traditional congeneric model employs a single scaling parameter because it links only the true scores of two measures. The geometric framework developed later generalizes this transformation by allowing separate scaling of population means, true-score dispersion, and error dispersion.

Equation (1) describes transformations of individual true scores using a single scaling parameter. However, when measurement transformations are examined at the population level, distinct transformations may act on population means, true-score dispersions, and error dispersions. To represent these possibilities separately, the broader framework developed below introduces three scaling parameters $\gamma_1, \gamma_2, \gamma_3$. The parameter $\gamma_1$ governs the transformation of population means, $\gamma_2$ governs the transformation of population true-score standard deviations, and $\gamma_3$ governs the transformation of population error-score standard deviations. Together, these parameters define the transformation structure from which the geometric analysis that follows is constructed.

**Transformation Matrices**

Classical measurement theory specifies how scores from different instruments relate to one another through linear transformations of true scores and error scores and, consequently, of their population standard deviations. To study measurement transformations geometrically, it is useful to collect the principal population characteristics of a measure into a single measurement-state vector (McDonald, 1999). These transformations of true and error variability can be expressed using upper-triangular matrices acting on a measurement state vector, $m_A$, of population true score means, true scores SDs and error SDs,

$$m_A = \begin{pmatrix} \mu(\tau_A) \\ \sigma(\tau_A) \\ \sigma_\epsilon(Y_A) \\ 1 \end{pmatrix}, \tag{3}$$

where $\mu(\tau_A)$ is the population mean true score on measure A, $\sigma(\tau_A)$ denotes the population true score SD on measure $A$, $\sigma_\epsilon(Y_A)$ the population error standard deviation of scores on measure $A$, and the constant 1 allows affine shifts. These three coordinates are chosen because they represent the principal location and variability characteristics of a measurement distribution and will later serve as the coordinates of the geometric representation. The measurement state vector represents a population distribution of scores on measure A (McDonald, 1999). In classical theory, since the expected value of the error is 0, then,

$$\mu(Y_A) = \mu(\tau_A),$$

and so the population mean observed score is equal to the population mean true score (Lord & Novick, 1968, theorem 2.7.2, p. 37). Because the expected error is zero, the population mean observed score equals the population mean true score, so no information is lost by representing only the latter.

Each classical equivalence class, or form of parallelism (Haertel, 2006), corresponds to a particular form of upper-triangular transformation matrix. Parallel forms have identical true-score and error standard deviations, hence equal reliability coefficients. The transformation of measurement state vector $m_A$ (eqn. 3) to measurement state vector $m_B$,

$$m_B = \begin{pmatrix} \mu(\tau_B) \\ \sigma(\tau_B) \\ \sigma_\epsilon(Y_B) \\ 1 \end{pmatrix}$$

is therefore the identity,

$$\Gamma_{\text{parallel}} = \begin{pmatrix} 1 & 0 & 0 & 0 \\ 0 & 1 & 0 & 0 \\ 0 & 0 & 1 & 0 \\ 0 & 0 & 0 & 1 \end{pmatrix}. \quad (4)$$

Tau-equivalent measures have equal true scores, hence equal true score variances, but may have equal or different error standard deviations. For simplicity, and because the subsequent geometric development focuses on equal-scaling transformations, attention is restricted to the equal-error-SD subcase. Although parallel and tau-equivalent measures differ conceptually through their assumptions about error variances, under the equal-error-SD restriction adopted here both have the identity transformation on the measurement-state vector (Green & Carroll, 1976; Lord & Novick, 1968; Mulaik, 2010; Nunnally & Bernstein, 1994).

Essentially tau-equivalent measures have true scores that differ only by an additive constant, $\omega$, with $\omega \in \mathbb{R}$, so the population true score SD is the same as for tau-equivalent and parallel measures, and if error SDs are equal, the reliabilities are equal, and the transformation is,

$$\Gamma_{\text{essentially tau-equivalent}} = \begin{pmatrix} 1 & 0 & 0 & \omega \\ 0 & 1 & 0 & 0 \\ 0 & 0 & 1 & 0 \\ 0 & 0 & 0 & 1 \end{pmatrix} \tag{5}$$

so the action on a measurement state vector will be given by eqn. (6) below,

$$\Gamma_{\text{essentially tau-equivalent}} m_B = \begin{pmatrix} 1 & 0 & 0 & \omega \\ 0 & 1 & 0 & 0 \\ 0 & 0 & 1 & 0 \\ 0 & 0 & 0 & 1 \end{pmatrix} \begin{pmatrix} \mu(\tau_A) \\ \sigma(\tau_A) \\ \sigma_\epsilon(Y_A) \\ 1 \end{pmatrix} = \begin{pmatrix} \mu(\tau_A) + \omega \\ \sigma(\tau_A) \\ \sigma_\epsilon(Y_A) \\ 1 \end{pmatrix},$$

(Haertel, 2006; Green & Carroll, 1976; Krantz et al., 1972; Lord & Novick, 1968; Savov, 2017).

Congeneric measures allow both a scaling of true scores and an additive shift, along with a potentially different scaling of error SDs (Haertel, 2006; Jöreskog, 1971). A particularly important special case, as will be seen below, is a symmetric congeneric transformation, meaning that the true-score population mean, true score SD, and error-score standard deviation are scaled by the same factor $\gamma$, so the transformation becomes,

$$\Gamma_{\text{sym}} = \begin{pmatrix} \gamma & 0 & 0 & \omega \\ 0 & \gamma & 0 & 0 \\ 0 & 0 & \gamma & 0 \\ 0 & 0 & 0 & 1 \end{pmatrix}, \tag{7}$$

with $\gamma$ a real number greater than 0, or symbolically $\gamma \in \mathbb{R}_{>0}$, and $\omega$ a real number $\mathbb{R}$. In this special case, the action on the measurement state vector is,

$$m_B = \begin{pmatrix} \gamma & 0 & 0 & \omega \\ 0 & \gamma & 0 & 0 \\ 0 & 0 & \gamma & 0 \\ 0 & 0 & 0 & 1 \end{pmatrix} \begin{pmatrix} \mu(\tau_A) \\ \sigma(\tau_A) \\ \sigma_\epsilon(Y_A) \\ 1 \end{pmatrix} = \begin{pmatrix} \gamma\mu(\tau_A) + \omega \\ \gamma\sigma(\tau_A) \\ \gamma\sigma_\epsilon(Y_A) \\ 1 \end{pmatrix}. \tag{8}$$

Thus, the parallel and tau-equivalent cases can be seen as special cases of (8) in which $\gamma = 1$ and $\omega = 0$; and essentially tau-equivalent cases can be seen as special cases of (8) in which $\gamma = 1$ and $\omega \in \mathbb{R}$ (Lord & Novick, 1968). In each of these cases it can be shown the reliability coefficients are equal.

Matrix (7) is a special case of a more general transformation,

$$\Gamma(\gamma_1, \gamma_2, \gamma_3, \omega) = \begin{pmatrix} \gamma_1 & 0 & 0 & \omega \\ 0 & \gamma_2 & 0 & 0 \\ 0 & 0 & \gamma_3 & 0 \\ 0 & 0 & 0 & 1 \end{pmatrix}, \tag{8a}$$

with $\gamma_1, \gamma_2, \gamma_3 \in \mathbb{R}_{>0}$, and $\omega \in \mathbb{R}$, which will be considered in detail below. Equation (8a) represents the most general scaling of the measurement-state vector coordinates. Unlike the symmetric, equal-scaling case expressed by (7), the parameters $\gamma_1, \gamma_2, \gamma_3$ are allowed to vary independently, allowing population means, true-score dispersions, and error dispersions to transform at different rates. The equal scaling case, $\gamma_1 = \gamma_2 = \gamma_3$, is of particular interest because it preserves reliability and the ratio structure of the variance components while leading naturally to the geometric results developed below. *Matrix (7) is a special and important case of (8a) in which* $\gamma_1 = \gamma_2 = \gamma_3 = \gamma$.

The transformation rules for transformation matrices (4), (5), (7) and (8a) form a special family of upper-triangular matrices. They are closed under matrix multiplication, as shown by, the multiplication of two matrices of the form (8a), $\Gamma(\gamma_1, \gamma_2, \gamma_3, \omega)$ and $\Gamma'(\gamma_1, \gamma_2, \gamma_3, \omega)(\gamma'_1, \gamma'_2, \gamma'_3, \omega)$,

$$\Gamma(\gamma_1, \gamma_2, \gamma_3, \omega)\Gamma'(\gamma_1, \gamma_2, \gamma_3, \omega)(\gamma'_1, \gamma'_2, \gamma'_3, \omega) = \begin{pmatrix} \gamma_1\gamma'_1 & 0 & 0 & \gamma_1\omega' + \omega \\ 0 & \gamma_2\gamma'_2 & 0 & 0 \\ 0 & 0 & \gamma_3\gamma'_3 & 0 \\ 0 & 0 & 0 & 1 \end{pmatrix}, \quad (9)$$

which shows the group product law as,

$$\Gamma(\gamma_1, \gamma_2, \gamma_3, \omega)\Gamma(\gamma'_1, \gamma'_2, \gamma'_3, \omega) = \Gamma(\gamma_1\gamma'_1, \gamma_2\gamma'_2, \gamma_3\gamma'_3, \gamma_1\omega' + \omega).$$

The inverse of (8a) is,

$$\Gamma^{-1}(\gamma_1, \gamma_2, \gamma_3, \omega) = \begin{pmatrix} \frac{1}{\gamma_1} & 0 & 0 & -\frac{\omega}{\gamma_1} \\ 0 & \frac{1}{\gamma_2} & 0 & 0 \\ 0 & 0 & \frac{1}{\gamma_3} & 0 \\ 0 & 0 & 0 & 1 \end{pmatrix};$$

the identity is,

$$\Gamma(1,1,1,0) = \begin{pmatrix} 1 & 0 & 0 & 0 \\ 0 & 1 & 0 & 0 \\ 0 & 0 & 1 & 0 \\ 0 & 0 & 0 & 1 \end{pmatrix};$$

and matrix multiplication is associative. Because the matrices depend smoothly on the parameters $(\gamma_1, \gamma_2, \gamma_3, \omega)$, and because multiplication and inversion are smooth operations, $\Gamma(\gamma_1, \gamma_2, \gamma_3, \omega)$ forms a four-dimensional Lie group. (Nugent, 2024, 2026; Stillwell, 2000). The Lie-group structure provides a continuous symmetry framework in which infinitesimal and global measurement transformations can be studied within a unified geometric setting. This group in which $\gamma_1 = \gamma_2 = \gamma_3 = \gamma$, is referred to as the $\Gamma_{sym}(\gamma, \omega)$ group in what follows. The equal-scaling subgroup $\Gamma_{sym}(\gamma, \omega)$ will play a notable role in what follows because as shown below, its action preserves the faithful Heisenberg representation (Stillwell, 2000).

**The Heisenberg Group**

The Heisenberg group is a noncommutative matrix Lie group and, like all Lie groups, is a differentiable manifold. The standard matrix representation of the Heisenberg group is,

$$H(a, b, c) = \begin{pmatrix} 1 & a & c \\ 0 & 1 & b \\ 0 & 0 & 1 \end{pmatrix}, \tag{10}$$

where, $a, b, c \in \mathbb{R}$, and 1s are on the diagonal. To facilitate direct comparison with the 4×4 measurement-transformation matrices introduced above, a faithful 4×4 representation of the Heisenberg group will be used throughout the remainder of the paper. This representation is,

$$H(a, b, c) = \begin{pmatrix} 1 & a & c & 0 \\ 0 & 1 & b & 0 \\ 0 & 0 & 1 & 0 \\ 0 & 0 & 0 & 1 \end{pmatrix}, \tag{11}$$

again where, $a, b, c \in \mathbb{R}$. This 4×4 representation is faithful to the standard Heisenberg group (11) (Kolenkow, 2022). The family of Heisenberg matrices forms a Lie matrix group of 4×4 matrices under matrix multiplication. The Heisenberg group plays a significant role in areas including quantum mechanics, signal processing, harmonic analysis, and sub-Riemannian geometry. Of particular relevance here is that *the Heisenberg group is the simplest nontrivial nilpotent Lie group exhibiting noncommutativity*.

The Heisenberg group operation is matrix multiplication. For two Heisenberg group elements (i.e., Heisenberg matrices), $H(a,b,c)$ and $H'(a',b',c')$, their matrix product is given by equation (12) below,

$$H(a,b,c)H'(a',b',c') = \begin{pmatrix} 1 & a & c & 0 \\ 0 & 1 & b & 0 \\ 0 & 0 & 1 & 0 \\ 0 & 0 & 0 & 1 \end{pmatrix} \begin{pmatrix} 1 & a' & c' & 0 \\ 0 & 1 & b' & 0 \\ 0 & 0 & 1 & 0 \\ 0 & 0 & 0 & 1 \end{pmatrix} =$$

$$\begin{pmatrix} 1 & a+a' & c+c'+ab' & 0 \\ 0 & 1 & b+b' & 0 \\ 0 & 0 & 1 & 0 \\ 0 & 0 & 0 & 1 \end{pmatrix}, \qquad (12)$$

or in a more compact notation,

$$H(a,b,c)H'(a',b',c') = H(a+a', b+b', c+c'+ab'), \qquad (12a)$$

equations that express the *Heisenberg group law*. The additional term $ab'$ in the upper-right entry reflects the noncommutativity of the group. This noncommutativity is evident from the fact that,

$$H(a,b,c)\, H'(a',b',c') \neq H'(a',b',c')\, H(a,b,c),$$

whenever $ab' \neq a'b$. The noncommutativity is encoded in the vertical direction in the Heisenberg geometry, described further below. In later sections, the horizontal coordinates of the Heisenberg group will be related to true-score and error-score variability, suggesting a possible interpretation of the noncommutative structure in measurement-theoretic terms (Hall, 2015; Montgomery, 2006).

The parameter $c$ occupies a distinguished role because matrices of the form $H(0,0,c)$ commute with all elements of the group. These matrices constitute the center of the Heisenberg group and correspond to the vertical direction in Heisenberg geometry. Thus, while the parameters $a$ and $b$ determine the horizontal directions of the group, the parameter $c$ determines the central (vertical) direction. The inverse of a Heisenberg group element is,

$$H^{-1}(a,b,c) = \begin{pmatrix} 1 & -a & ab-c & 0 \\ 0 & 1 & -b & 0 \\ 0 & 0 & 1 & 0 \\ 0 & 0 & 0 & 1 \end{pmatrix} = H(-a,-b,ab-c).$$

The geometric structures of the Lie and Heisenberg groups are defined over a type of mathematical space known as a manifold (Binz & Pods, 2008; Capogna et al., 2007; Folland, 2025; Stillwell, 2000). A manifold is a space that locally resembles Euclidean space even when its global structure is more complicated. The Heisenberg group is a three-dimensional smooth manifold with global coordinates $(x, y, z)$. At each point $p = (x, y, z)$, the tangent space, symbolized as $T_pH$, consists of all possible instantaneous directions of motion through the point $p$. The horizontal distribution is spanned by two left-invariant vector fields, conventionally denoted $X$ and $Y$, whose commutator generates the vertical direction. These horizontal and vertical directions play a central role in the sub-Riemannian geometry of the Heisenberg group, which is described next. Because its noncommutativity is governed by a single commutator, the Heisenberg group provides a particularly transparent setting for studying whether classical measurement transformations preserve a noncommutative structure (Binz & Pods, 2008; Capogna et al., 2007; Folland, 2025; Hall, 2015; Stillwell, 2000). Consequently, if the measurement transformation group preserves the Heisenberg group under conjugation, it also preserves this underlying geometric structure.

**The Heisenberg Geometry**

The non-Euclidean geometry of the Heisenberg group consists of two levels: a horizontal plane spanned by two left-invariant vector fields, $X$ and $Y$, and a vertical direction generated by a third vector field, $Z$. This horizontal–vertical structure is the core of Heisenberg geometry. At each point $(x, y, z)$ of the Heisenberg manifold, the vector fields $X$ and $Y$ span the horizontal plane and generate all allowable directions of "motion" through linear combinations of the form $aX + bY$. Thus, at each point, the admissible directions form a two-dimensional subspace of the three-dimensional tangent space of the Heisenberg geometry.

Although the full Heisenberg geometry is *sub-Riemannian*, the horizontal distribution is equipped with a Euclidean inner product (with $X$ and $Y$ forming an orthonormal basis). The horizontal metric is therefore,

$$ds^2 = dx^2 + dy^2,$$

so the horizontal directions behave like ordinary orthogonal axes, and distances within the horizontal plane are computed exactly as in two-dimensional Euclidean space. Consequently, all non-Euclidean behavior of the Heisenberg geometry is confined to the vertical direction generated by the commutator

The full space spanned by $X$, $Y$, and $Z$, however, is *non-Euclidean*. The vertical direction $Z$ is not freely accessible: one cannot move directly in the $Z$-direction. Vertical motion is produced only through order-dependent combinations of horizontal motions. In the Heisenberg group, the vertical direction is generated by the Lie bracket, $[X, Y]$, of the horizontal vector fields. The commutator structure of the horizontal distribution is represented mathematically by,

$$[X, Y] = \frac{\partial}{\partial z} = Z. \quad (13)$$

Equation (13) expresses the Lie bracket (or commutator) of the horizontal vector fields. Geometrically, this commutator has a simple interpretation: move a small amount in the $X$-direction, then a small amount in the $Y$-direction, then undo the $X$-motion, and finally undo the $Y$-motion. The result is not zero—you do not end up where you started but rather end up displaced in the vertical direction. *This is the defining feature of the sub-Riemannian geometry of the Heisenberg group: horizontal motion is free, while vertical motion is generated only by the interaction of horizontal motions*. The explicit coordinate expressions for these vector fields are not needed for the present development; only their commutation structure is required. The vertical displacement is not directly accessible. Instead, it depends on both the order of the horizontal motions and the area enclosed by their path. This path dependence is a hallmark of the Heisenberg geometry and motivates the later investigation of whether analogous accumulations arise when true-score and error components interact through measurement transformations. In later sections, the horizontal coordinates will be associated with true-score and error-score variability, while the vertical coordinate will be shown to emerge from their interaction under measurement transformations (Agrachev et al., 2019; Binz & Pods, 2008; Boothby, 1986; Capogna et al., 2007; Folland, 2025; Hall, 2015; Magnani 2012; Montgomery, 2006; Semmes, 2023).

**Conjugation**

To understand how classical measurement transformations interact with the intrinsic uncertainty structure encoded by the Heisenberg group, we examine the conjugation of a Heisenberg element $H(a, b, c)$ by the general transformation (8a), $\Gamma(\gamma_1, \gamma_2, \gamma_{3,} \omega)$. Conjugation is the standard Lie-theoretic tool for determining whether one symmetry preserves the intrinsic structure of another. If conjugation leaves every element unchanged, the two groups commute; if it changes elements while remaining within the subgroup, it defines an automorphism; and if it moves elements outside the subgroup, the subgroup is not preserved (Hall, 2015; Knapp, 2002; Horn & Johnson, 2012; Stillwell, 2000).

The following theorem establishes the central mathematical result of the paper

***Theorem 1***: (Conjugation preserves the Heisenberg representation).

*Let*

$$\Gamma(\gamma_1, \gamma_2, \gamma_3, \omega)$$

*be the measurement transformation group defined by Equation (8a), and let*

$$H(a, b, c)$$

*be the faithful Heisenberg representation defined by Equation (11), with* $\gamma_1, \gamma_2, \gamma_3 \in \mathbb{R}_{>0}$ *and all* $\omega, a, b, c \in \mathbb{R}$*. Then conjugation by* $\Gamma(\gamma_1, \gamma_2, \gamma_3, \omega)$ *defines an automorphism of the Heisenberg group by,*

$$\Gamma(\gamma_1, \gamma_2, \gamma_3, \omega) H(a, b, c) \Gamma^{-1}(\gamma_1, \gamma_2, \gamma_3, \omega) = H\left(\frac{\gamma_1}{\gamma_2} a, \frac{\gamma_2}{\gamma_3} b, \frac{\gamma_1}{\gamma_3} c\right).$$

*Moreover, the central coordinate satisfies,*

$$scale(c) = scale(a) scale(b).$$

**Proof:**

A direct computation gives,

$\Gamma(\gamma_1, \gamma_2, \gamma_{3,}\omega)H(a,b,c)\Gamma^{-1}(\gamma_1, \gamma_2, \gamma_{3,}\omega) =$

$$\begin{pmatrix} \gamma_1 & 0 & 0 & \omega \\ 0 & \gamma_2 & 0 & 0 \\ 0 & 0 & \gamma_3 & 0 \\ 0 & 0 & 0 & 1 \end{pmatrix} \begin{pmatrix} 1 & a & b & 0 \\ 0 & 1 & c & 0 \\ 0 & 0 & 1 & 0 \\ 0 & 0 & 0 & 1 \end{pmatrix} \begin{pmatrix} \gamma_1^{-1} & 0 & 0 & -\frac{\omega}{\gamma} \\ 0 & \gamma_2^{-1} & 0 & 0 \\ 0 & 0 & \gamma_3^{-1} & 0 \\ 0 & 0 & 0 & 1 \end{pmatrix} = \begin{pmatrix} 1 & a\frac{\gamma_1}{\gamma_2} & c\frac{\gamma_1}{\gamma_3} & 0 \\ 0 & 1 & b\frac{\gamma_2}{\gamma_3} & 0 \\ 0 & 0 & 1 & 0 \\ 0 & 0 & 0 & 1 \end{pmatrix}, \quad (13)$$

or in briefer notation,

$$\Gamma(\gamma_1, \gamma_2, \gamma_{3,}\omega)H(a,b,c)\Gamma^{-1}(\gamma_1, \gamma_2, \gamma_{3,}\omega) = H\left(\frac{\gamma_1}{\gamma_2}a, \frac{\gamma_2}{\gamma_3}b, \frac{\gamma_1}{\gamma_3}c\right). \quad (13a)$$

Now, $\frac{\gamma_1}{\gamma_2} \times \frac{\gamma_2}{\gamma_3} = \frac{\gamma_1}{\gamma_3}$,

or expressed differently,

$$scale(c) = scale(a)\ scale(b) = \left(\frac{\gamma_1}{\gamma_2}\right)\left(\frac{\gamma_2}{\gamma_3}\right) = \frac{\gamma_1}{\gamma_3}. \quad (14)$$

The transformed matrix in (13) is again a Heisenberg matrix, with its coordinates scaled according to (13a). Since the central coordinate scales as the product of the horizontal scalings as in (14), the defining commutator structure is preserved. Hence the conjugation action is an automorphism of the Heisenberg group. □

Thus, the scaling factor of the central coordinate equals the product of the scaling factors of the two horizontal coordinates. This is exactly the compatibility required by the Heisenberg commutator relation $[X, Y] = Z$, and therefore the conjugation action preserves the algebraic structure of the Heisenberg group.

**Corollary 1**: (Equal scaling case).

*In the equal scaling case,* $\gamma_1 = \gamma_2 = \gamma_3 = \gamma$, *the conjugation action is the identity on* $H$, $\Gamma_{sym}(\gamma, \omega)$ *lies in the centralizer of* $H(a, b, c)$, *and the faithful Heisenberg representation is invariant under conjugation by* $\Gamma_{sym}(\gamma, \omega)$.

**Proof**:

In the symmetry case in which, $\gamma_1 = \gamma_2 = \gamma_3 = \gamma$, the equal scaling case, let,

$$\Gamma_{sym}(\gamma, \omega) = \begin{pmatrix} \gamma & 0 & 0 & \omega \\ 0 & \gamma & 0 & 0 \\ 0 & 0 & \gamma & 0 \\ 0 & 0 & 0 & 1 \end{pmatrix}. \quad (15)$$

The results of the conjugation then reduce to,

$$\Gamma_{sym}(\gamma, \omega)\mathrm{H}(a, b, c)\Gamma_{sym}^{-1}(\gamma, \omega) = \begin{pmatrix} 1 & a & c & 0 \\ 0 & 1 & b & 0 \\ 0 & 0 & 1 & 0 \\ 0 & 0 & 0 & 1 \end{pmatrix}, \quad (16)$$

or in Heisenberg notation,

$$\Gamma_{sym}(\gamma, \omega)\mathrm{H}(a, b, c)\Gamma_{sym}^{-1}(\gamma, \omega) = H(a, b, c). \quad (16a)$$

Multiplying both sides of (16) on the right by $H^{-1}(a, b, c)$, gives the commutator,

$$[\Gamma_{sym}(\gamma, \omega)H(a, b, c)\Gamma_{sym}^{-1}(\gamma, \omega)]H^{-1}(a, b, c) = \begin{pmatrix} 1 & 0 & 0 & 0 \\ 0 & 1 & 0 & 0 \\ 0 & 0 & 1 & 0 \\ 0 & 0 & 0 & 1 \end{pmatrix} = I, \quad (16b)$$

so, the commutator is,

$$\left[\Gamma_{sym}(\gamma, \omega), H(a, b, c)\right] = I, \quad (17)$$

indicating that $\Gamma_{sym}(\gamma, \omega)$ and $H(a, b, c)$ commute. □

Therefore, the conjugation action is trivial, $\Gamma_{\text{sym}}(\gamma, \omega)$ lies in the centralizer of $H$, and the intrinsic Heisenberg geometry is invariant under changes of measurement frame generated by $\Gamma_{\text{sym}}(\gamma, \omega)$ (Hall, 2015; Knapp, 2002; Horn & Johnson, 2012; Stillwell, 2000).

**Proposition**: (Broken symmetry).

*If the equal-scaling symmetry fails, conjugation is nontrivial and the commutator is given by* $H\left(\left(\frac{\gamma_1}{\gamma_2}-1\right)a, \left(\frac{\gamma_2}{\gamma_3}-1\right)b, \left(\frac{\gamma_1}{\gamma_3}-1\right)c + \left(1-\frac{\gamma_1}{\gamma_2}\right)ab\right)$.

**Proof**:

If, $\gamma_1 \neq \gamma_2 \neq \gamma_3$, the commutator is,

$$[\Gamma(\gamma_1,\gamma_2,\gamma_3,\omega)H(a,b,c)\Gamma^{-1}(\gamma_1,\gamma_2,\gamma_3,\omega)]H^{-1}(a,b,c) =$$

$$\begin{pmatrix} 1 & a\frac{\gamma_1}{\gamma_2} & c\frac{\gamma_1}{\gamma_3} & 0 \\ 0 & 1 & b\frac{\gamma_2}{\gamma_3} & 0 \\ 0 & 0 & 1 & 0 \\ 0 & 0 & 0 & 1 \end{pmatrix} \begin{pmatrix} 1 & -a & ab-c & 0 \\ 0 & 1 & -b & 0 \\ 0 & 0 & 1 & 0 \\ 0 & 0 & 0 & 1 \end{pmatrix} =$$

$$\begin{pmatrix} 1 & \left(\frac{\gamma_1}{\gamma_2}-1\right)a & \left[\left(\frac{\gamma_1}{\gamma_3}-1\right)c + \left(1-\frac{\gamma_1}{\gamma_2}\right)ab\right] & 0 \\ 0 & 1 & \left(\frac{\gamma_2}{\gamma_3}-1\right)b & 0 \\ 0 & 0 & 1 & 0 \\ 0 & 0 & 0 & 1 \end{pmatrix}, \tag{18}$$

or in simpler notation,

$$[\Gamma(\gamma_1,\gamma_2,\gamma_3,\omega)H(a,b,c)\Gamma^{-1}(\gamma_1,\gamma_2,\gamma_3,\omega)]H^{-1}(a,b,c)$$

$$= H\left(\left(\frac{\gamma_1}{\gamma_2}-1\right)a, \left(\frac{\gamma_2}{\gamma_3}-1\right)b, \left(\frac{\gamma_1}{\gamma_3}-1\right)c + \left(1-\frac{\gamma_1}{\gamma_2}\right)ab\right). \square \tag{18a}$$

Thus, when the symmetry $\gamma_1 = \gamma_2 = \gamma_3$ breaks, $\Gamma(\gamma_1,\gamma_2,\gamma_3,\omega)$ and $H(a,b,c)$ fail to commute.

Note that if $\gamma_1 = \gamma_2$ the cross-product term, ab, drops out of the commutator. If $\gamma_1 = \gamma_3$, the contribution of $c$ to the central coordinate vanishes, although the $ab$ contribution generally remains. In the fully symmetric case, $\gamma_1 = \gamma_2 = \gamma_3 = \gamma$, then,

$$\frac{\gamma_1}{\gamma_2} = \frac{\gamma_2}{\gamma_3} = \frac{\gamma_1}{\gamma_3} = \frac{\gamma}{\gamma} = 1,$$

so in this symmetry case, as found above,

$$\left[\Gamma_{sym}(\gamma, \omega), H(a, b, c)\right] = H(0,0,0) = I,$$

so the two groups commute.

**Lemma**: *The intersection of* $\Gamma(\gamma_1, \gamma_2, \gamma_3, \omega)$ *with* $H(a, b, c)$ *is the identity.*

**Proof**:

Suppose $H(a, b, c) = \Gamma(\gamma_1, \gamma_2, \gamma_3, \omega)$. Comparing matrix entries gives,

$$H(a, b, c) \cap \Gamma(\gamma_1, \gamma_2, \gamma_3, \omega) \rightarrow H(a, b, c) = \Gamma(\gamma_1, \gamma_2, \gamma_3, \omega),$$

so the intersection implies,

$$\begin{pmatrix} \gamma_1 & 0 & 0 & \omega \\ 0 & \gamma_2 & 0 & 0 \\ 0 & 0 & \gamma_3 & 0 \\ 0 & 0 & 0 & 1 \end{pmatrix} = \begin{pmatrix} 1 & a & c & 0 \\ 0 & 1 & b & 0 \\ 0 & 0 & 1 & 0 \\ 0 & 0 & 0 & 1 \end{pmatrix},$$

which holds only when $\gamma_1 = \gamma_2 = \gamma_3 = 1$, and $a = b = c = 0$, and $\omega = 0$. Thus, the intersection is the identity (Hall, 2015; Knapp, 2002; Horn & Johnson, 2012) □ .

**A Direct Product Structure**

These results reveal a strong symmetry and geometric compatibility: the transformations $\Gamma_{sym}(\gamma, \omega)$ preserve the Heisenberg structure exactly. The two geometric structures coexist independently: every element of $H(a, b, c)$ commutes with every element of $\Gamma_{sym}(\gamma, \omega)$. This strong symmetry is itself a result of the symmetry, $\gamma_1 = \gamma_2 = \gamma_3 = \gamma$, which shows the scalings by $\Gamma_{sym}(\gamma, \omega)$ of the population mean, the true score SD, and the error SD are the same. As eqn. (18) shows, the strong symmetry and geometric compatibility breaks when the symmetry $\gamma_1 = \gamma_2 = \gamma_3$ fails to hold.

**Corollary 2**: *The groups $H(a, b, c)$ and $\Gamma_{sym}(\gamma, \omega)$ form a direct product.*

**Proof**:

Since every element of $H$ commutes with every element of $\Gamma_{\mathrm{sym}}$, and since $H \cap \Gamma_{\mathrm{sym}} = \{I\}$, the subgroup generated by $H(a, b, c)$ and $\Gamma_{\mathrm{sym}}(\gamma, \omega)$ is the internal direct product $H \times \Gamma_{\mathrm{sym}}$. In Lie-group terms, the two groups form a direct product:

$$H(a, b, c) \times \Gamma_{sym}(\gamma, \omega) = \begin{pmatrix} 1 & a & c & 0 & 0 & 0 & 0 & 0 \\ 0 & 1 & b & 0 & 0 & 0 & 0 & 0 \\ 0 & 0 & 1 & 0 & 0 & 0 & 0 & 0 \\ 0 & 0 & 0 & 1 & 0 & 0 & 0 & 0 \\ 0 & 0 & 0 & 0 & \gamma & 0 & 0 & \omega \\ 0 & 0 & 0 & 0 & 0 & \gamma & 0 & 0 \\ 0 & 0 & 0 & 0 & 0 & 0 & \gamma & 0 \\ 0 & 0 & 0 & 0 & 0 & 0 & 0 & 1 \end{pmatrix}, \quad (20)$$

with each layer acting independently. □

The group multiplication law for (20) is, in Heisenberg notation equation (21),

$$(a, b, c, \gamma, \omega)\,(a', b', c', \gamma', \omega') = (a + a', b + b', c + c' + ab', \gamma\gamma', \gamma\omega' + \omega).$$

This expresses the full symmetry uncovered by the conjugation, commutator, and intersection results. This two-layer direct product itself is a Lie group and a manifold.

Therefore, in the equal-scaling case $\gamma_1 = \gamma_2 = \gamma_3$, the subgroup generated by $H(a, b, c)$ and $\Gamma_{sym}(\gamma, \omega)$ is the internal direct product $H \times \Gamma_{sym}$. Equivalently, every Heisenberg element is fixed under conjugation by $\Gamma_{sym}(\gamma, \omega)$, so the symmetric transformation group lies in the centralizer of the Heisenberg group and preserves the intrinsic Heisenberg geometry exactly. *By symmetry argument, if a particular measurement procedure provides a valid geometric representation of a Heisenberg type phenomenon, then so will the representation transformed by* $\Gamma_{sym}(\gamma, \omega)$ (Hall, 2015; Knapp, 2002; Horn & Johnson, 2012; Rosen, 1995; Van Fraassen, 1989).

## Discussion

The principal mathematical result of this paper is that the measurement transformation group $\Gamma(\gamma_1, \gamma_2, \gamma_3, \omega)$ acts by automorphisms on the Heisenberg group and that, under the equal-scaling symmetry $\gamma_1 = \gamma_2 = \gamma_3$, the conjugation action becomes trivial. In this symmetry case the Heisenberg subgroup is fixed under conjugation, the symmetric transformation group lies in its centralizer, and the subgroup generated by the two groups is their internal direct product. From a measurement perspective, the equal-scaling symmetry identifies a class of transformations that preserve the intrinsic noncommutative geometry represented by the Heisenberg group. When the symmetry is broken, the nontrivial commutator quantifies departures from that geometry.

The equal-scaling symmetry, $\gamma_1 = \gamma_2 = \gamma_3$, differentiates two classes of measurement transformations: those that meet the symmetry condition preserve the intrinsic Heisenberg geometry, whereas those that violate the symmetry produce a nontrivial commutator that measures departures from the symmetry. In geometric terms, the Heisenberg representation remains unchanged when transformations of means, true-score dispersions, and error dispersions remain proportionally aligned by the symmetry, $\gamma_1 = \gamma_2 = \gamma_3$. When this alignment fails, the resulting commutator indicates the extent of the distortion introduced into the representation. Taken together, these results establish a geometric framework in which classical measurement theory provides the transformation symmetry, the Heisenberg group provides a canonical representation of noncommutative phenomena, and conjugation characterizes the compatibility between them.

The commutator given in eqn. (18) - (18a) acquires a geometric interpretation. It provides a geometric measure of departures from the equal-scaling symmetry. If the symmetry $\gamma_1 = \gamma_2 = \gamma_3$ holds, then the commutator in (18) vanishes and reduces to the identity. So, when $\gamma_1 = \gamma_2 = \gamma_3$, $\Gamma_{\text{sym}}(\gamma, \omega)$ and $H(a, b, c)$ commute. As the scaling parameters depart from the equal-scaling condition, the commutator departs from the identity. The resulting nontrivial commutator provides a quantitative measure of the degree to which the equal-scaling symmetry has been broken.

The central coordinate, c, is not introduced as an independent bias parameter. It arises naturally from the geometry of the Heisenberg group. The horizontal motions in the XY plane give rise to the central term, c, and movement in the Z direction by way of the commutator, $[X, Y] = Z$. The "motions" in the XY plane are changes in true score (X axis) and error (Y axis) SDs. In the Heisenberg geometry the induced motion in the Z direction represents interactions between the true score SD and the error SD; this is the noncommutativity. This arises as a consequence of the Heisenberg geometry. Note that from the conjugation results (13a) and (14) that the scaling of the central coordinate is the product of the scalings of the a and b coordinates.

Again, this demonstrates that the central coordinate arises as the geometric interaction between true score SDs and error SDs in the horizontal plane spanned by the X and Y directions in the Heisenberg geometry, as opposed to an ad hoc addition to represent order effects or other forms of bias. The theorem shows that the Heisenberg group is compatible with the measurement transformation geometry precisely because its commutator structure is preserved under conjugation.

It is also noteworthy that the reciprocal conjugation exhibits the same symmetry. Under the equal-scaling condition,

$$H(a,b,c)\Gamma_{sym}(\gamma,\omega)H^{-1}(a,b,c) = \Gamma_{sym}(\gamma,\omega),$$

so conjugation by a Heisenberg element leaves every element of the symmetric transformation subgroup unchanged. When the equal-scaling symmetry is broken, however, the reverse conjugation acquires additional off-diagonal shear terms proportional to the differences among the scaling parameters. Thus, the reciprocal conjugation exhibits the same symmetry-breaking behavior as the conjugation of the Heisenberg group by the transformation group. Together these results show that, under the equal-scaling symmetry, neither $H(a,b,c)$ nor $\Gamma_{sym}(\gamma,\omega)$ alters the matrix representation of the other under conjugation. The two mutual conjugation actions are therefore reciprocal: each subgroup is fixed under conjugation by the other, reflecting the direct-product structure $H \times \Gamma_{sym}$. When the symmetry is broken, both conjugation actions become nontrivial. Thus, under the equal-scaling symmetry, each subgroup fixes the other under conjugation, providing a reciprocal characterization of the direct-product structure $H \times \Gamma_{sym}$.

The present work establishes a bridge between two mathematical structures that have traditionally been studied independently. Classical measurement theory contributes the transformation symmetry, while the Heisenberg group contributes the simplest noncommutative geometry capable of representing interaction-generated effects. The conjugation theorem shows how these structures fit together.

The contribution of the present work is not the introduction of the Heisenberg group itself, nor a reformulation of classical measurement theory. Rather, it is the demonstration that the transformation structure already implicit in classical measurement theory possesses a symmetry that acts compatibly with the canonical noncommutative geometry of the Heisenberg group. This compatibility is expressed mathematically through the conjugation and direct-product results developed above. The symmetry structure already present in classical measurement theory is compatible with the canonical geometry of the simplest noncommutative Lie group.

The present analysis is primarily theoretical. Its implications for applied measurement remain to be investigated empirically. The foregoing suggests a number of lines for future research. First, requirements for test equating include: the forms to be equated must measure the exact same construct, have equal reliabilities, and it should be a matter of indifference to test takers which form is used, among others (Kolen & Brennan, 2014; Holland & Dorens, 2006). These requirements are naturally interpreted as symmetry conditions. Thus, one area of future research suggested is investigating equating by way of symmetry considerations and more specifically the $\Gamma_{sym}(\gamma, \omega)$ group of transformations. Another area concerns meta-analysis. The $\Gamma_{sym}(\gamma, \omega)$ group of transformations imply a number of measurement invariants, including reliability, standardized regression coefficients, and the population standardized mean difference effect size (Nugent, 2024, 2026). This suggests that the comparability of scores from different measures in meta-analysis follows from the symmetry principle if the scores from the different measures are related by the $\Gamma_{sym}(\gamma, \omega)$ group.

The present results also suggest that constructs in social, behavioral, and medical science fields exhibiting order effects, path dependence, response-context effects, or interaction-dependent measurement behavior may benefit from representation within a Heisenberg framework. This may address a gap in current measurement theory. Finally, the forgoing suggests that the use of Lie group theory holds promise for further investigations and theoretical development in measurement and applied measurement theory.

## Conclusion

The present analysis demonstrates that classical measurement transformations possess a previously unrecognized Lie-group structure and that these transformations act by automorphisms on a faithful Heisenberg representation. The conjugation theorem identifies a distinguished equal-scaling symmetry under which the Heisenberg representation is preserved exactly, the conjugation action becomes trivial, and the subgroup generated by the transformation group and the Heisenberg group reduces to an internal direct product. These mathematical results establish a geometric bridge between classical measurement theory and the simplest noncommutative Lie group. They further suggest that noncommutative geometric representations may provide a useful framework for studying measurement systems in which interaction effects, order dependence, and path-dependent behavior are important. More broadly, echoing recommendations by Nugent (2024, 2026), the results suggest that Lie group methods may offer a productive mathematical framework for extending classical measurement theory, with potential applications to problems such as test equating, measurement invariance, and meta-analysis.

References

Agrachev, A., Barilari, D., & Boscain, U. (2019). *A comprehensive introduction to sub-Riemannian geometry*. Cambridge University Press.

Allen, M. J., & Yen, W. M. (1979). *Introduction to measurement theory.* Brooks/Cole.

Atmanspacher, H., & Filk, T. (2010). A proposed test of temporal nonlocality in bistable perception. *Journal of Mathematical Psychology, 54*, 314-321. https://doi.org/10.1016/j.jmp.2009.12.001

Atmanspacher, H., & Romer, H. (2018). Order effects in sequential measurements of non-commuting psychological observables. arXiv:1201.4685v2. https://arxiv.org/pdf/1201.4685. Retrieved June 22, 2026.

Binz, E., & Pods, S. (2008). *The Geometry of Heisenberg Groups: With Applications in Signal Theory, Optics, Quantization, and Field Quantization*. Mathematical Surveys and Monographs, 151. American Mathematical Society.

Boothby, W. M. (1986). *An introduction to differentiable manifolds and Riemannian geometry* (2nd ed.). Academic Press.

Borsboom, D. (2005). *Measuring the mind: Conceptual issues in contemporary psychometrics.* Cambridge University Press.

Busemeyer, J., & Bruza, P. (2024). *Quantum Models of Cognition and Decision*. Cambridge University Press.

Capogna, L., Pauls, S., & Danielli, D. (2007). The Heisenberg Group and Sub-Riemannian Geometry. In: Tyson, J.T. (eds) *An Introduction to the Heisenberg Group and the Sub-Riemannian Isoperimetric Problem,* pp. 11 – 37. Progress in Mathematics, vol 259. Birkhäuser Basel. https://doi.org/10.1007/978-3-7643-8133-2_2. Retrieved: June 20, 2026.

Conte, E., Khrennikov, A. Y., Todarello, O., Federici, A., & Zbilut, J. P. (2008). On the existence of quantum wave function and quantum interference effects in mental states: An experimental confirmation during perception and cognition in humans. *arXiv preprint arXiv:0807.4547*. Retrieved June 20, 2026.

Dzhafarov, E. N., Kujala, J. V., & Cervantes, V. (2016). Contextuality-by-default: A brief overview of ideas, concepts, and terminology. https://arxiv.org/abs/1504.00530. Retrieved July 9, 2026.

Folland, G. B. (2025). *The Heisenberg Group: A Survey*. American Mathematical Society.

Golan. J. (2012). The linear algebra a beginning graduate student ought to know (3rd ed). Springer.

Ghiselli, E., Campbell, J., & Zedeck, S. (1981). *Measurement theory for the behavioral sciences*. W.H. Freeman and Company.

Green, P. & Carroll, J.D. (1976). *Mathematical tools for applied multivariate analysis*. (Student edition). Academic Press.

Guttman, L. (1945). A basis for analyzing test–retest reliability. *Psychometrika, 10*(4), 255–282.

https://doi.org/10.1007/BF02288892

Haertel, E. (2006). Reliability. In R. Brennan (Ed.), *Educational Measurement* (4th ed.), pp. 65 – 110. American Council on Education/Praeger.

Hall, B. C. (2015). *Lie groups, lie algebras, and representations: An elementary introduction* (2nd ed.). Springer.

Holland, P. W., & Dorans, N. J. (2006). Linking and equating. In R. L. Brennan (Ed.), *Educational measurement* (4th ed., pp. 187–220). Praeger.

Horn, R. A., & Johnson, C. R. (2012). *Matrix Analysis* (2nd ed.). Cambridge University Press. https://doi.org/10.1017/cbo9781139020411

Jöreskog, K. G. (1971). Statistical analysis of sets of congeneric tests. *Psychometrika, 36*(2), 109–133. https://doi.org/10.1007/BF02291393

Khrennikov, A. (2010). *Ubiquitous quantum structure: From psychology to finance.* Springer.

Kolenkow, R. (2022). *An introduction to groups and their representations*. Cambridge University Press.

Kolen, M. J., & Brennan, R. L. (2014). *Test equating, scaling, and linking: Methods and practices* (3rd ed.). Springer.

Kolenkow, R. (2022). An introduction to groups and their matrices for science students. Cambridge University Press.

Knapp, A. (2002). *Lie groups: Beyond an introduction*. Birkhäuser.

Krantz, D. H., Luce, R. D., Suppes, P., & Tversky, A. (1971). *Foundations of measurement: Vol. I. Additive and polynomial representations*. Academic Press.

Lord, F. M., & Novick, M. R. (1970). *Statistical theories of mental test scores.* Addison-Wesley.

Magnani, V. (2012). An introduction to the geometries of Heisenberg groups. https://people.dm.unipi.it/magnani/Corsi/Cincinnati.pdf. Retrieved June 20, 2026.

McDonald, R. P. (1999). *Test theory: A unified treatment*. Lawrence Erlbaum Associates.

Montgomery, R. (2006). *A tour of sub-Riemannian geometries, their geodesics and applications*. American Mathematical Society.

Morgan, P. (2022). Classical and Quantum Measurement Theory. https://arxiv.org/abs/2201.04667. Retrieved: July 1, 2026.

Mulaik, S. A. (2010). *Foundations of factor analysis* (2nd ed.). CRC Press.

Nugent, W. R. (2026). Extending a Matrix Lie Group Model of Measurement Symmetries. *Symmetry*, *18*(2), 361. https://doi.org/10.3390/sym18020361

Nugent, W. R. (2024). A Matrix Lie Group Formulation of Measurement Theory: Symmetries of Classical Measurement Theory. *Measurement: Interdisciplinary Research and Perspectives*, *22*(3), 297–314. https://doi.org/10.1080/15366367.2023.2258483

Nunnally, J., & Bernstein, I. (1994). Psychometric theory (3rd ed.). McGaw-Hill.

Pothos, E. M., & Busemeyer, J. R. (2013). Can quantum probability provide a new direction for

cognitive modeling. *Behavioral and Brain Sciences, 36*(3), 255–327. https://doi.org/10.1017/S0140525X12001525

Rosen, J. (1995). Symmetry in Science. Springer-Verlag.

Savov, I. (2017). *No bullshit guide to linear algebra* (2nd ed.). Minireference Co.

Schwarz, N., & Sudman, P. (1992). *Context effects in social and psychological research*. Springer-Verlag.

Semmes, S. (2023). An Introduction to Heisenberg Groups in Analysis and Geometry. *Notices of the AMS*, *50*(6), 640-646. https://www.ams.org/notices/200306/fea-semmes.pdf. Retrieved: June 20, 2026.

Stillwell, J. (2000). *Naïve Lie Theory*. Springer.

Traub, R. (1997). Classical test theory in historical perspective. *Educational Measurement: Issues and Practice*, 16(4), 8 - 14. https://doi.org/10.1111/j.1745-3992.1997.tb00603.x

Trueblood, J.S. & Busemeyer, J.R. (2011), A Quantum Probability Account of Order Effects in Inference. *Cognitive Science*, *35*: 1518-1552. https://doi.org/10.1111/j.1551-6709.2011.01197.x

Van Fraassen, B. (1989). *Laws and Symmetry*. Oxford.

Wang, Z., Solloway, T., Shiffrin, R. M., & Busemeyer, J. R. (2014). Context effects produced by question orders reveal quantum nature of human judgments. *Proceedings of the National Academy of Sciences of the United States of America*, *111*(26), 9431–9436. https://doi.org/10.1073/pnas.1407756111